\documentclass{article}
\usepackage[T1]{fontenc}
\usepackage[latin9]{inputenc}
\usepackage{amsmath}
\usepackage{amssymb}
\usepackage{graphicx}
\usepackage{esint}
\usepackage[authoryear]{natbib}

\makeatletter

\providecommand{\tabularnewline}{\\}

\usepackage{times}\usepackage{amsfonts}\usepackage{epsf}
\def\x {\ifmmode \times\else $\times$\fi}\def\to {\ifmmode \tau_0\else $\tau_0$\fi}\def\Ea {\ifmmode \bar{a}\else $\bar{a}$\fi}\def\Eb {\ifmmode \bar{b}\else $\bar{b}$\fi}\def\dr {\ifmmode \delta{r}\else $\delta{r}$\fi}\def\nmin {\ifmmode n_{\rm min}\else $n_{\rm min}$\fi}\def\nobs {\ifmmode n_{\rm obs}\else $n_{\rm obs}$\fi}\def\fout {\ifmmode f_{\rm out}\else $f_{\rm out}$\fi}\def\serr {\ifmmode \sigma_{\rm err}\else $\sigma_{\rm err}$\fi}\def\rbin {\ifmmode r_{\rm bin}\else $r_{\rm bin}$\fi}\def\Epa {\ifmmode E^+_a\else $E^+_a$\fi}\def\Ema {\ifmmode E^-_a\else $E^-_a$\fi}\def\Epb {\ifmmode E^+_b\else $E^+_b$\fi}\def\Emb {\ifmmode E^-_b\else $E^-_b$\fi}\def\Vpa {\ifmmode V^+_a\else $V^+_a$\fi}\def\Vma {\ifmmode V^-_a\else $V^-_a$\fi}\def\Vpb {\ifmmode V^+_b\else $V^+_b$\fi}\def\Vmb {\ifmmode V^-_b\else $V^-_b$\fi}\def\Cpmab {\ifmmode C^\pm_{a,b}\else $C^\pm_{a,b}$\fi}\def\mdtt {\ifmmode \overline{\Delta t}_\tau \else $\overline{\Delta t}_\tau$\fi}\def\mdt {\ifmmode \overline{\Delta t} \else $\overline{\Delta t}$\fi}

\makeatother

\begin{document}

\title{Improved AGN light curve analysis with the $z$-transformed discrete
correlation function}

\author{Tal Alexander\\
 {\footnotesize Department of Particle Physics \& Astrophysics}\\
{\footnotesize Faculty of Physics}\\
{\footnotesize Weizmann Institute of Science}\\
 {\footnotesize POB 26, Rehovot 76100, Israel}\\
 {\footnotesize Email: tal.alexander@weizmann.ac.il}}
\maketitle
\begin{abstract}
The cross-correlation function (CCF) is commonly employed in the study
of AGN, where it is used to probe the structure of the broad line
region by line reverberation, to study the continuum emission mechanism
by correlating multi-waveband light curves and to seek correlations
between the variability and other AGN properties. The $z$-transformed
discrete correlation function (ZDCF) is a new method for estimating
the CCF of sparse, unevenly sampled light curves. Unlike the commonly
used interpolation method, it does not assume that the light curves
are smooth and it does provide errors on its estimates. The ZDCF corrects
several biases of the discrete correlation function method of Edelson
\& Krolik (1988) by using equal population binning and Fisher's $z$-transform.
These lead to a more robust and powerful method of estimating the
CCF of sparse light curves of as few as 12 points. Two examples of
light curve analysis with the ZDCF are presented. 1) The ZDCF estimate
of the auto-correlation function is used to uncover a correlation
between AGN magnitude and variability time scale in a small simulated
sample of very sparse and irregularly sampled light curves. 2) A maximum
likelihood function for the ZDCF peak location is used to estimate
the time-lag between two light curves.\textsc{ fortran 77} and \textsc{fortran
}95 code implementations of the ZDCF and the maximum likelihood peak
location algorithms are freely available (see \texttt{\footnotesize http://www.weizmann.ac.il/weizsites/tal/research/software/}). 
\end{abstract}

\section{Introduction}

The problem of analysing sparse, unevenly sampled light curves is
frequently encountered in the study of AGN, notably in line reverberation
mapping and in multi-wavelength variability studies. In many cases,
despite great observational efforts, the light curves can not be reliably
analysed by Fourier inversion or other inversion methods \citep{Maoz}.
There is, to date, no satisfactory alternative but to carry the analysis
in the time domain. A widely used tool for extracting information
from such light curves is the cross-correlation function (CCF) \citet{BM}
\begin{equation}
{\rm CCF(\tau)}=\frac{E[(a(t)-E_{a})(b(t+\tau)-E_{b})]}{\sqrt{V_{a}V_{b}}}\,,\label{e:CCF}
\end{equation}
 where $a$ and $b$ are the two light curves, $\tau$ the time-lag,
$E$ the expectation value and $V$ the variance over the light curve.

The CCF of AGN continuum and emission line light curves is used in
line reverberation mapping to estimate the size and geometry of the
broad line region \citet[and references therein]{Peterson}. The CCF
between different AGN continuum wave bands is used for studying the
continuum emission mechanism by looking for causal connection between
the various wavebands (e.g. Korista et al. 1995). Other uses of the
CCF include the determination of the Hubble constant from the time
lag between variations of the multiple images of a macro-lensed QSO
(e.g. Vanderriest et al. 1989) and the linear reconstruction of discretely
sampled light curves, whose input is the time-lag dependence of the
auto-correlation function (ACF) \citet{RP}.

\begin{figure}[t]
 \includegraphics[width=0.9\columnwidth]{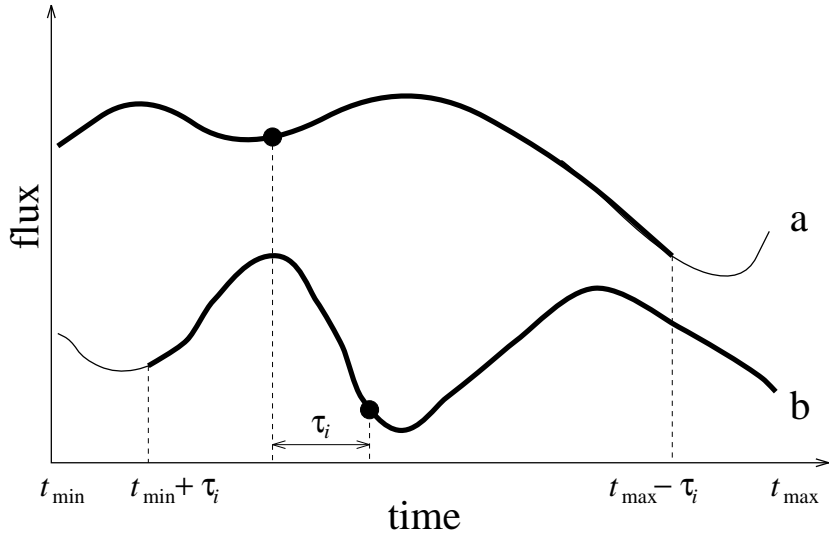} \caption{\label{f:ranges} The two observed points, with time lag $\tau_{i}$,
sample the correlation at this lag between the bold intervals of the
two continuous segments of the light curves.}
\end{figure}

In practice, finite segments of the light curves are sampled discretely
with measurement errors, at unevenly spaced intervals. The expectation
values and variances of the light curves are unknown and must also
be estimated from the observations. Three conditions are implicitly
assumed in the process of estimating the correlation function: 1)
statistical stationarity, 2) the ergodic assumption (i.e. that the
ensemble average of all the light curves, which are statistically
similar to the observed one, is equivalent to the time average over
a single infinite light curve) and 3) random sampling of the light
curves. If these are satisfied, then the data at hand can indeed yield
an estimate for the population correlation between signals that are
statistically similar to the observed light curves. The observed data
are the result of a two level sampling process. First, Nature `draws'
a light curve from the continuum of possible light curves. Second,
the observations sample the continuum of points in a finite segment
of this given light curve (See Fig.~\ref{f:ranges}). Conditions
(1) and (2) relate to first sampling level. They must usually be {\em
assumed}, since they cannot be inferred from the observed data. The
ZDCF (like the two other CCF estimators that are described below)
deals only with the second level of sampling, where the parent distribution
is the continuous distribution of the points in the given finite segments.

There are in general two approaches for dealing with the uneven sampling:
interpolation and the discrete correlation function (DCF). The interpolation
method of \citet{GP} calculates the CCF by averaging the cross-correlation
obtained by pairing the observed $a(t_{i})$ with the interpolated
value $b(t_{i}-\tau)$, and that obtained by pairing the observed
$b(t_{j})$ with the interpolated $a(t_{j}+\tau)$. The DCF method
\citet[EK]{EK} initially uses the observed points to calculate 
\begin{equation}
{\rm DCF}_{ij}=\frac{(a_{i}-\Ea^{\prime})(b_{j}-\Eb^{\prime})}{s_{a}^{\prime}s_{b}^{\prime}}\,\label{e:udcf}
\end{equation}
 where $a_{i}$, $b_{j}$ are the observed fluxes at times $t_{i}$
and $t_{j}$, respectively, and $\Ea^{\prime}$, $\Eb^{\prime}$,
$s_{a}^{\prime2}$ and $s_{b}^{\prime2}$ are the sample means and
variances over the entire light curves. Subsequently, $\{{\rm DCF}_{ij}\}$
are binned by their associated time-lag, $\tau_{ij}=t_{i}-t_{j}$,
into equal width bins, $\tau\pm\delta\tau$. The bin average is used
to estimate CCF$(\tau)$ and the bin's standard deviation to estimate
the error.

The two methods were compared by several authors \citet{RSC,WP} in
the context of AGN variability studies, and the general conclusion
seems to be that the interpolation method is equal or superior to
the DCF in most cases. Nevertheless, interpolation has several obvious
drawbacks. Adding interpolated points between those actually observed
amounts to inventing data or assuming that the light curve varies
smoothly. The interpolation method does not give error estimates on
the reconstructed CCF, which are necessary for model fitting. The
DCF concept is a more cautious approach to reconstructing the CCF.
However, the original DCF method has several problems, which are discussed
in detail below. These problems can have adverse effects on the DCF
in the realistic case of poorly and unevenly sampled light curves.

This work does not attempt to deal with the many potential pitfalls
in the statistical interpretation of the CCF and its significance,
which may arise from incorrect assumptions of stationarity or ergodicity
or from an incorrect measurement error model (e.g. Box \& Newbold
1971). The use of the CCF in AGN light curve analysis is dictated
by the present-day quality of the data and the nature of current AGN
research. In this given situation, it is important to try and develop
more reliable methods for estimating the CCF. 
The validity of the underlying assumptions in the case of AGN will
be ultimately justified if the accumulated results from many observation
campaigns converge into a coherent physical picture.

The approach taken here is empirical, in that analytical arguments
and approximate assumptions are used only as motivation for the proposed
improvements, whose final test is by simulations with AGN-like light
curves. Care is taken to verify that in cases where the assumptions
do not hold, the CCF estimate is conservative rather than deceptively
significant. Several changes in the original DCF method are suggested,
the major ones being the use of the $z$-transform and equal population
binning. These result in the $z$-transformed discrete correlation
function (ZDCF), which is an improved, more robust method for reconstructing
the CCF under realistically unfavourable conditions.

The ZDCF method is described and contrasted with the DCF in section~\ref{s:calc}.
Their performance is compared by simulations in section~\ref{s:results}.
Section~\ref{s:ex} presents two sample applications of the ZDCF.
The usefulness of the ZDCF in extracting information from a small
number of observations is demonstrated in section~\ref{s:ex1}, where
it is used to find a correlation between the variability time scale
and magnitude in a small sample of simulated AGN light curves. Section~\ref{s:ex2}
presents a maximum likelihood method for estimating the time-lag between
two light curves. The results are discussed and summarised in section~\ref{s:discuss}.

\section{The ${\bf z}$-transformed Discrete Correlation Function}

\label{s:calc}

\subsection{Estimating the correlation with the $z$-transform}

Let $n$ be the number of $\{a_{i},b_{i}\}$ pairs in a given time-lag
bin. CCF($\tau$) is estimated by the correlation coefficient 
\begin{equation}
r=\frac{\sum_{i}^{n}(a_{i}-\Ea)(b_{i}-\Eb)/(n-1)}{s_{a}s_{b}}\,.\label{e:rbin}
\end{equation}
 where $\Ea$, $\Eb$ are the estimators of the bin averages, and
$s_{a}$, $s_{b}$ the estimators of the standard deviations 
\begin{equation}
s_{a}^{2}=\frac{1}{n-1}\sum_{i}^{n}(a_{i}-\bar{a})^{2}\,,
\end{equation}
 with $s_{b}^{2}$ is similarly defined. Note, in contrast, that the
DCF is not normalized correctly since the moments of the full light
curves, which appear in ${\rm DCF}_{ij}$ (equation~\ref{e:udcf}),
are used to normalize the individual bins. The correct normalization,
as well as ensuring that $|r|\le1$, reduces the errors when the light
curves are non-stationary by limiting the summation to those points
that actually contribute to the bin \citet{WP}. The sampling distribution
of $r$ is known to be highly skewed and far from normal and therefore
estimating its sampling error by the sample variance $s_{r}$ can
be very inaccurate. When $a$ and $b$ are drawn from the bivariate
normal distribution it is possible to transform $r$ into an approximately
normally distributed random variable, Fisher's $z$ (e.g. Kendall
\& Stuart 1969, p. 390, Kendall \& Stuart 1973, p. 486 and references
therein). Defining 
\begin{equation}
z=\frac{1}{2}\log\left(\frac{1+r}{1-r}\right)\,,\,\,\,\zeta=\frac{1}{2}\log\left(\frac{1+\rho}{1-\rho}\right)\,,\,\,\, r=\tanh z\,,\label{e:z}
\end{equation}
 the mean and variance of $z$ are approximately equal to 
\begin{eqnarray}
\bar{z} & = & \zeta+\frac{\rho}{2(n-1)}\times\nonumber \\
 &  & \left[1+\frac{5+\rho^{2}}{4(n-1)}+\frac{11+2\rho^{2}+3\rho^{4}}{8(n-1)^{2}}+\cdots\right]\,,
\end{eqnarray}
 and 
\begin{equation}
s_{z}^{2}=\frac{1}{n-1}\left[1+\frac{4-\rho^{2}}{2(n-1)}+\frac{22-6\rho^{2}-3\rho^{4}}{6(n-1)^{2}}+\cdots\right]\,,\label{e:sz}
\end{equation}
 where $\rho$ is the unknown population correlation coefficient of
the bin. In order to estimate $\bar{z}$ and $s_{z}$, the {\em
ansatz} $\rho=r$ is assumed. Transforming back to $r$, the interval
corresponding to the normal $\pm1\sigma$ error interval can be estimated
by 
\begin{equation}
\dr_{\pm}=|\tanh(\bar{z}(r)\pm s_{z}(r))-r|\,.\label{e:dcferr}
\end{equation}
 The validity of this {\em ansatz} was verified by simulations
with normal bivariate distributions of different correlation coefficients
and $n$ = 11. The results show that the probability of $\rho$ lying
outside the empirical interval $r\pm\delta r_{\pm}$ is $1.03$ to
$1.08$ times higher than the normal value of 0.3174, implying a slight
under-estimation of the errors. 
 Unlike the $r\pm{s_{r}}$ error interval of the DCF, the $z$-transformed
error interval satisfies $|r\pm\dr_{\pm}|\le1$. As $|r|$ tends to
1 it becomes both asymmetric around $r$ and smaller than $r\pm{s_{r}}$,
thus improving the error estimates in the physically interesting extrema
of the CCF. The normality of $z$ is known to hold down to $n=11$
for an exact normal bivariate distribution or for $\rho=0$ and otherwise
tends asymptotically with $n$ towards normality. As will be discussed
below, the dependence of the $z$-transform on the shape of the parental
distribution plays an important role in determining the properties
and limitations of the ZDCF. The effects of deviations from binormality,
and especially from mesokurtosis, can be significant \citet{Gayen}.
While the resulting bias in $\bar{z}$ is moderate and tends to zero
as $n$ increases, the bias in $s_{z}$ may be non-negligible and
its behaviour as $n$ increases varies depending on the underlying
bivariate distribution. Nevertheless, the $n=11$ lower limit still
holds as long as the deviations from binormality are moderate. There
exist also further refinements of the $z$ transform, $z^{*}$ and
$z^{**}$ \citet{Hotelling}. Their effect on the ZDCF was checked
by simulations, similar to those described in section \ref{s:results}
below and the improvement in the results was found to be negligible
and not worth the added calculational complexity.

\subsection{The binning method}

\label{s:bin}

The ZDCF binning method differs from that of EK in both the binning
criterion and the treatment of interdependent pairs in the bin. The
ZDCF bins by equal population and as a result the bins are not equal
in time-lag width. Each bin contains at least $\nmin=11$ points,
(the minimum for a meaningful statistical interpretation) and does
not contain interdependent pairs, which are discarded. Consequently,
light curves of less then 12 observations cannot be analysed by this
method.

\subsubsection{Statistical properties}

The main source of bias in estimating the bin's correlation coefficient
is the existence of inner correlations within the sets $\{a_{i}\}$
and $\{b_{i}\}$. The mean time between the observations in a bin
associated with time-lag $\tau$ is 
\begin{equation}
\mdtt=\frac{T-\tau}{n_{{\rm min}}}\label{e:mdt}
\end{equation}
 where $T$ is the duration of the two observed light curves (for
simplicity, it will be assumed from this point on that the two light
curves were observed over the same period). If the light curve is
significantly auto-correlated over a coherence time scale $\to$,
then when $\mdtt<\to$, the sampling is no longer random since consecutive
points in the bin are not independent%
\footnote{The coherence time scale which is relevant here is that of finite
segments of length $T$ and not that of infinite light curves, which
may be larger. This distinction is relevant, for example, for light
curves with power-law power spectra, where $\to$ increases with $T$
as the lower frequency, higher amplitude variations become more dominant.%
}. $\mdtt$ should not be confused with the mean time between observations,
$\mdt=T/(\nobs-1)$. $\mdtt$ is not a function of the number of observations,
$\nobs$, and therefore equal population binning automatically stabilises
the results to changes in $\nobs$.

It is instructive to discuss the effect of auto-correlation on the
$z$ transform in terms of its marginal parental distributions. The
irregularly spaced observation times, $\{t_{i}\}$, can be viewed
as sampling the distribution of a random variable $t$. The observed
points are therefore distributed as functions of the random variable
$t$, namely $a(t)$ and $b(t)$. Since these distributions are not
necessarily normal, the $z$ transform may be biased, even when the
signals are not auto-correlated. The bias caused by auto-correlation
has a distinctive signature which can be demonstrated by studying
smooth light curves in the limit of short $T$. In this case the coherence
time scales of both light curves are of the order of $T$ itself,
the observed segments are almost linear and the distributions of $\{a_{i}\}$
and $\{b_{i}\}$ are approximately linear functions of the distribution
of $t$. Suppose further that $t$ is uniformly distributed, so that
${kurt}(t)=-1.2$. Since the kurtosis is invariant under linear transformations
of the random variable, the kurtosis of $a$ and $b$ will also be
-1.2. Negative kurtosis in $a$ or $b$ can contribute to significant
over-estimation of $s_{z}$ \citet[eqs. 81--84]{Gayen}, since 
\begin{eqnarray}
s_{z,\,{\rm bias}}^{2} & = & s_{z,\,{\rm true}}^{2}-\nonumber \\
 &  & \frac{(n-3-\rho^{2})}{(n-1)^{2}}\frac{\rho^{2}({kurt}(a)+{kurt}(b))}{4(1-\rho^{2})^{2}}+\cdots
\end{eqnarray}
 This qualitative explanation of the tendency of the ZDCF to over-estimate
the errors of auto-correlated signals is supported by the simulations
in section~\ref{s:results}. The DCF is also known to suffer from
this bias.

Astronomical observations are often clustered on daily or seasonal
timescales. One simple model for this distribution is that of $N$
periods, each divided into two sub-periods. At the first sub-period,
of length $T_{1}$, the object can be observed, and the observing
times are uniformly distributed. At the second sub-period, of length
$T_{2}$, the object cannot be observed. It can be shown that the
kurtosis of this distribution is also negative, $kurt=-1.2F(N,T_{2}/T_{1})$,
where the function $F$ is bracketed between 1 and 5/3 and approaches
1 very rapidly with increasing $N$. Such sampling patterns will therefore
also lead to over-estimation of the sampling errors.

The robustness of $\bar{z}$ and $s_{z}$ to deviations from the assumed
binormality of the parental distributions determines their usefulness
in analysing auto-correlated light curves. The nature and magnitude
of these deviations depend on the ratio $\mdtt/\to$ and on the specific
statistical properties of the light curve. The simulations below show
that in the case of AGN-like data sets, these deviations are still
small enough for the $z$-transform to be of practical use.

Another source of bias is the occurrence of interdependent pairs,
such as $\{a_{i},b_{j}\}$ and $\{a_{i},b_{k}\}$, in the same bin.
If, for example, $a$ and $b$ are fully correlated at time-lag $\tau$,
$\, a(t)\propto b(t+\tau)\,$, then $r(\tau)$ will be biased towards
zero because the repeated occurrences of the point $a_{i}$ are equivalent
to substituting the true $a$ by a constant light curve. Although
the frequency of these multiple occurrences can be lowered by decreasing
the bin width, this is inconsistent with the requirement that the
bin contain at least 11 points. The ZDCF addresses this problem directly
by discarding the interdependent pairs. This is to be contrasted with
the suggestion of EK that the effect of these pairs can be taken into
account by heuristically increasing the DCF error estimates beyond
the actual scatter in the bin.

The binned correlation coefficient associates with the mean bin lag,
$\bar{\tau}$, an estimate of the bin average of the correlation coefficient.
This is not the same as estimating $\rho(\bar{\tau})$. The approximation
involved in using the binned average and its interpretation are discussed
in appendix~\ref{s:appA}.

\subsubsection{The binning algorithm}

The binning is implemented by ordering all the possible pairs $\{a_{i},b_{j}\}$
by their associated time-lag $\tau=t_{i}-t_{j}$. The ordered list
is then divided, bin by bin, into bins of $\nmin$ pairs. In the process
of adding pairs to a bin, a new pair whose $a$ or $b$ points have
previously appeared in that bin, is discarded. The artificial separation
of pairs of very close time-lags into adjacent bins is prevented by
defining a small parameter $\epsilon$ so that a new pair with time-lag
$\tau_{i+1}$ will be added to the bin as long as $\tau_{i+1}-\tau_{i}<\epsilon$,
even if $\nmin$ is exceeded. Most of the information in the ZDCF
is at the time lags where the overlap between the two light curves
is large (cf. Fig.~\ref{f:ccf}). It is therefore desirable that
the binning algorithm minimize the bin widths at these lags. This
can be achieved by starting the allocation of pairs to bins at the
lag where the pairs are densest. For the auto-correlation function,
this means that the binning proceeds from $\tau=0$ up to $\tau_{{\rm max}}$.
For the CCF, the binning proceeds from the median $\tau$ up to $\tau_{{\rm max}}$
and then from the median $\tau$ down to $\tau_{{\rm min}}$.

The allocation of pairs to bins depends on the order in which the
binning is carried out and on the choice of the interdependent pairs
which are discarded. While this does not affect the average statistical
properties of this method, different choices may lead to different
ZDCF results for a given poorly sampled light curve. This ambiguity
can be resolved when the ZDCF points are used for fitting a model
CCF. In this case the fit score can be averaged over the possible
binning choices.

\subsection{Measurement errors}

\label{s:error}

The measured light curves $a$, $b$ are the sum of the true light
curves $x$, $y$ and random noises $n_{a}$, $n_{b}$ 
\begin{equation}
a(t_{i})=x(t_{i})+n_{a}(t_{i})\,\,\,\,\,\,\,,\,\,\,\,\,\,\, b(t_{j})=y(t_{j})+n_{b}(t_{j})\,.
\end{equation}
 As the measurement errors propagate into the CCF in a complicated,
non-linear manner, it is suggested that their effect be estimated
by Monte Carlo simulations. This is performed by assuming a distribution
for the measurement errors (usually the normal distribution), adding
a randomly drawn error value to each point and recalculating the ZDCF.
The Monte Carlo average of $z$ is then used to obtain $r$ and $\dr_{\pm}$.
The resulting ZDCF is conservative in that it estimates the mean CCF
that is {\em consistent\/} with the observations and $\sqrt{2}$
times the quoted measurement errors. The over-estimated errors are
an unavoidable consequence of having no model for the light curve
but the observed points themselves. The errors on the ZDCF points
are the sampling errors and should not be interpreted as estimating
the deviation from the CCF of the true signals due to the measurement
errors. For example, the ZDCF of very noisy light curves tends to
zero irrespective of the correlation between the true signals. It
should be emphasized that the data points are not weighted by their
errors. It is therefore recommended that points with atypically high
measurement errors (relative to the mean error) be judiciously discarded
from the light curve before the ZDCF is applied.

This approach to estimating the effects of the measurement errors
is safer than that suggested by EK, which diverges in the limit of
a small number of observations. This is discussed in more detail in
appendix~\ref{s:appB}

\section{Results}

\label{s:results}

\begin{figure*}
\begin{tabular}{c}
\includegraphics[bb=12bp 12bp 576bp 482bp,clip,width=0.75\columnwidth]{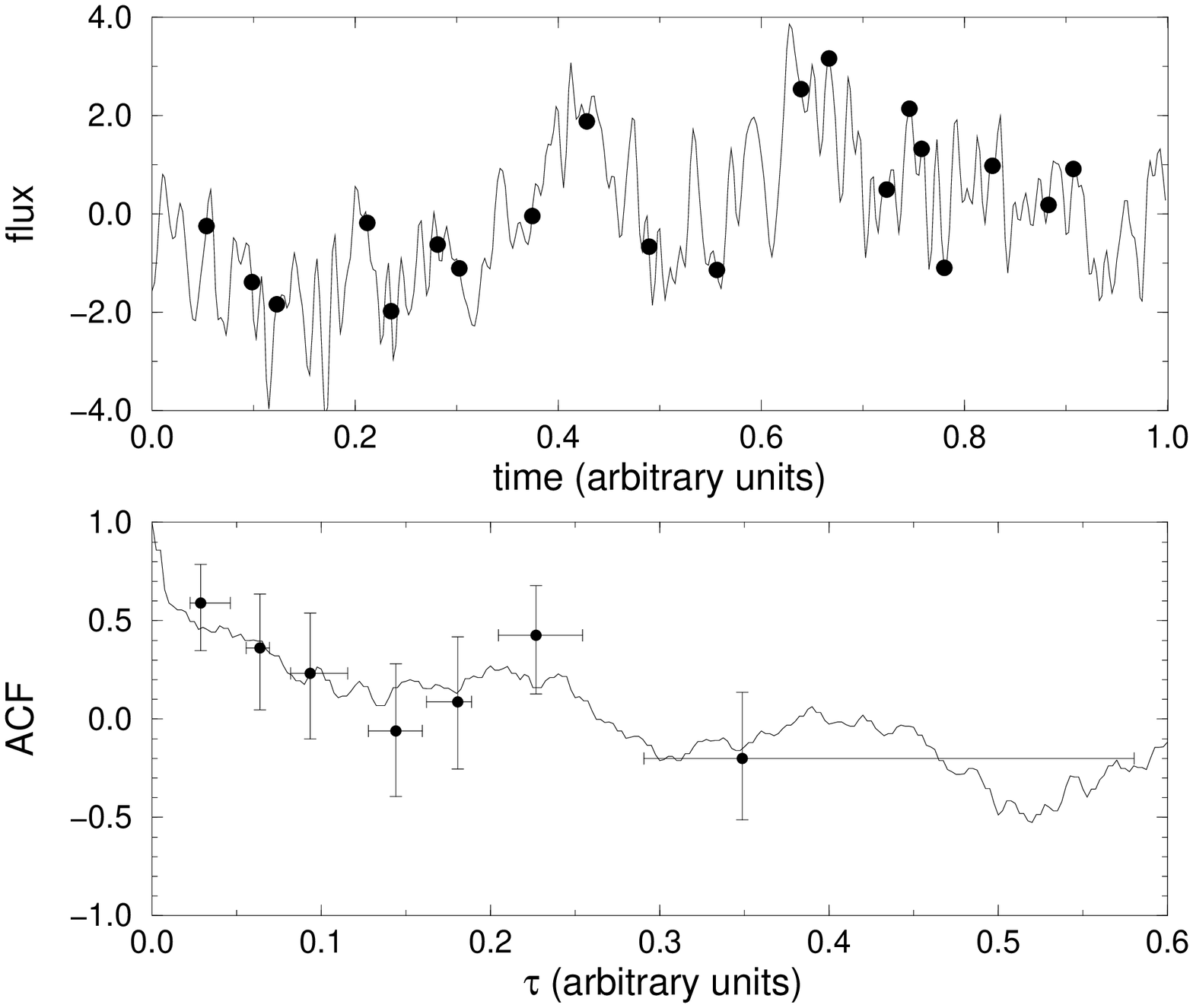}\tabularnewline
\includegraphics[width=0.75\columnwidth]{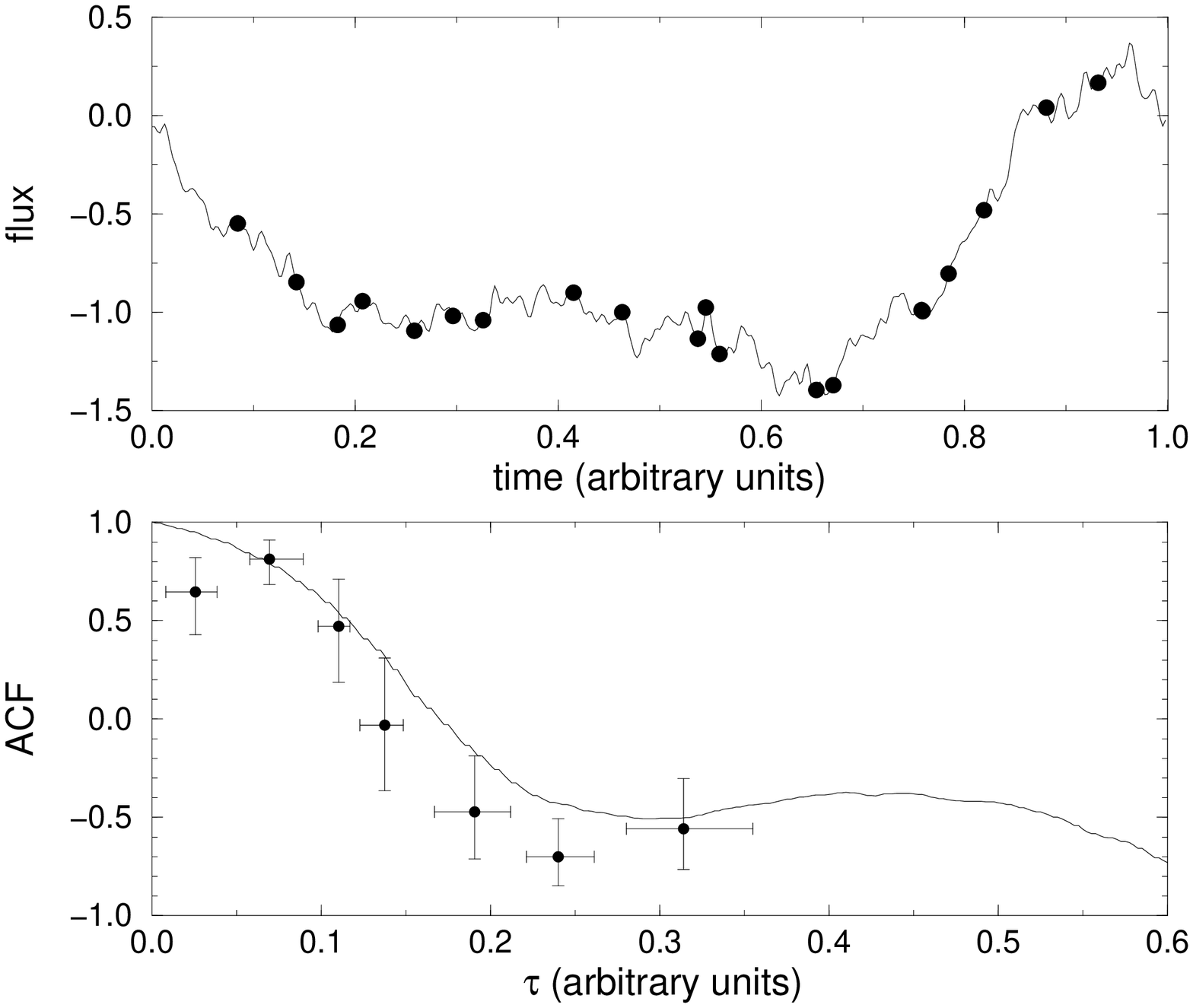}\tabularnewline
\end{tabular} \caption{\label{f:LC}Two examples of simulated light curves and their ZDCF.
The time differences between each successive pair of the 20 observations
(black dots) are drawn from the uniform distribution. No measurement
errors added. Top 2 panels: flicker-noise light curve ($P(\nu)\propto1/\nu$)
and its ACF. Bottom 2 panels: soft spectrum light curve ($P(\nu)\propto1/\nu^{2}$)
and its ACF. The simulated light curve is shown in solid line and
the discrete observations in dots. The true ACF is shown in solid
line and the auto-correlation function calculated by the ZDCF method
is shown as points with error bars.}
\end{figure*}

\begin{figure*}
\begin{tabular}{c}
\includegraphics[width=0.75\columnwidth]{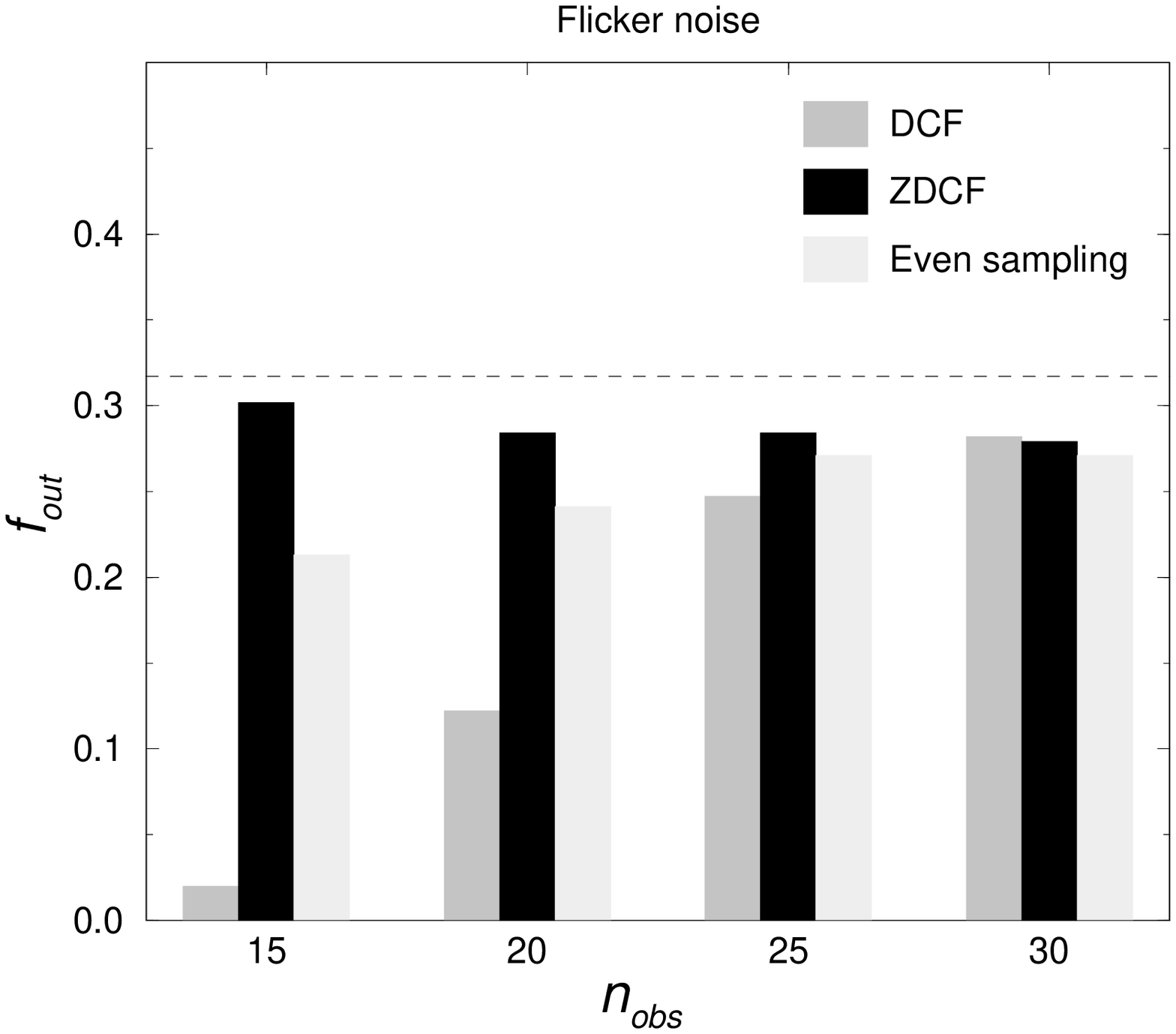}\tabularnewline
\includegraphics[width=0.75\columnwidth]{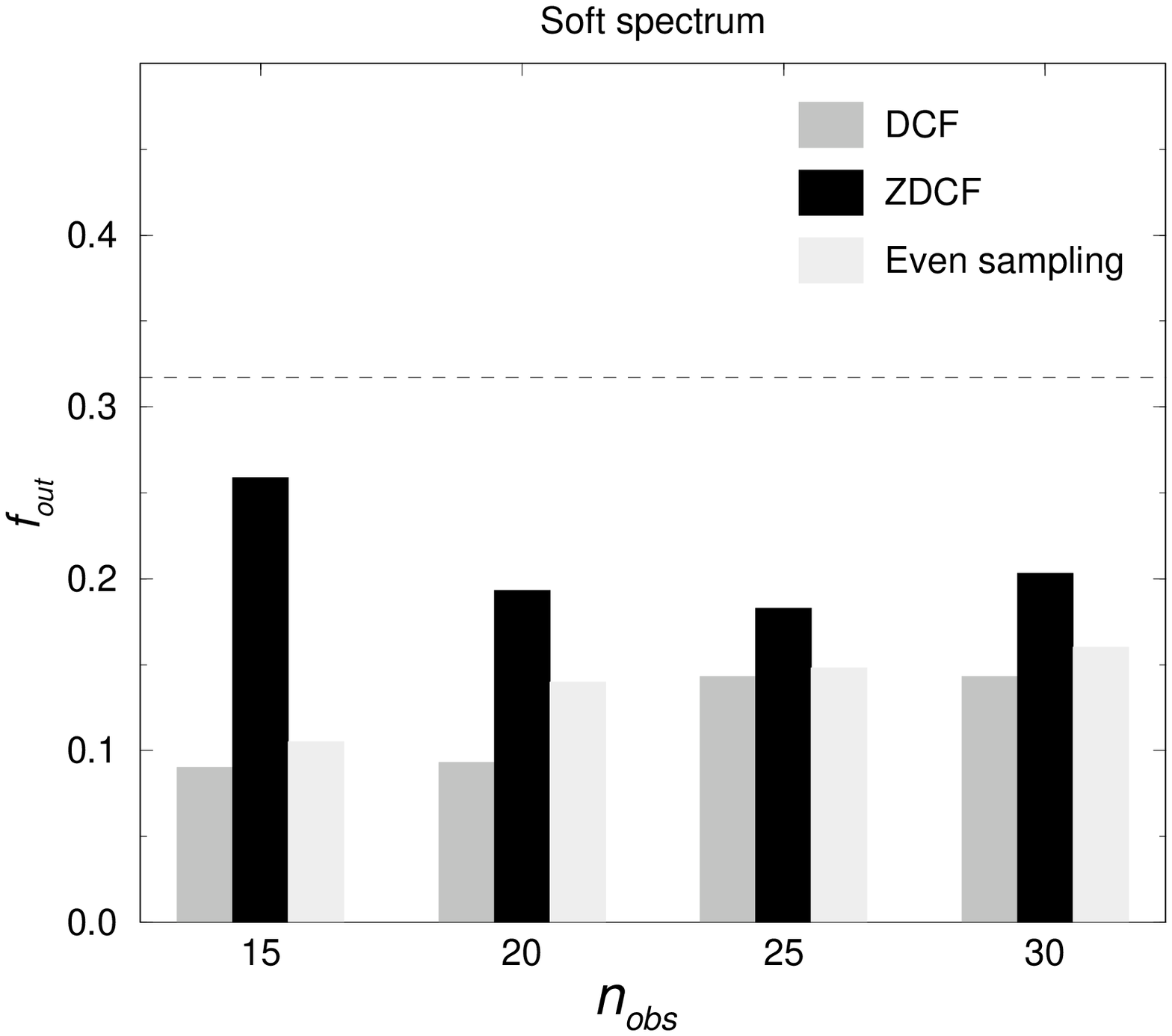}\tabularnewline
\end{tabular} \caption{\label{f:outliers}The fraction of outlying points, $\fout$ as function
of the number of simulated observations, $\nobs$. The results are
the average over 100 simulations. Top: flicker-noise light curves.
Bottom: soft, $P(\nu)\propto1/\nu^{2}$, light curves. The even sampled
bins were thinned out so as to have only 11 points per bin. No measurement
errors added.}
\end{figure*}

The properties of AGN power spectra are currently well known only
in the X-ray range, where the power spectrum can be approximated by
a power law, $P(\nu)\propto\nu^{-\alpha}$, with $1\lesssim\alpha\lesssim2$
\citet{GML}. In order to compare the performance of the DCF and ZDCF
on AGN-like light curves, the two methods were applied to two types
of simulated light curves (Fig~\ref{f:LC}): Noisy light curves with
a $P(\nu)\propto\nu^{-1}$ power spectrum (flicker noise) and softer
light curves with a $P(\nu)\propto\nu^{-2}$ power spectrum. Both
were generated by choosing random phases for frequencies in the range
$\nu_{{\rm min}}=1/4$, $\nu_{{\rm max}}=100$ at increments of $\Delta\nu=1/4$
(in arbitrary inverse time units). A light curve was then calculated
from $t=0$ to 1. 

The simulations consisted of drawing eight sets of 100 light curves,
one for each of the two power spectra and $\nobs=15,20,25$ and $30$.
For each of these light curves, the times of the $\nobs$ simulated
observations were randomly drawn so that the time difference between
two successive observations was uniformly distributed. The light curve
was then sampled at these random times and the discrete auto-correlation
function was calculated by both the DCF and ZDCF methods. In order
to have a common basis for comparing the two methods, the number of
bins used by the DCF was adjusted to equal the mean one used by the
ZDCF. The ZDCF was calculated with $\nmin=11$ which resulted in $\langle n_{{\rm bin}}\rangle=2,6,14$
and $22$ for $\nobs=15,20,25$ and $30$, respectively. In order
to isolate the possible effects of the binning procedure on the results,
eight additional sets of simulated light curves were similarly generated
and sampled at evenly spaced time intervals. The ZDCF was then calculated
after the resulting zero width bins were thinned down to 11 randomly
chosen points per bin to avoid possible biases due to the $z$ transform's
$n$ dependence. The fit of the ZDCF to the true ACF was evaluated
by the fraction of outlying bins, $\fout$, whose error interval does
not include $\rho_{{\rm bin}}$, which was calculated from the full
light curve for each bin. Similarly, for the DCF the comparison was
to $\bar{\rho}$, averaged over the interval $\tau_{i}\pm\delta\tau$.
For each of the two light curve types the average kurtosis of $a$
and $b$ and a $\chi^{2}$ fit of the marginal distributions to the
normal distribution were also calculated to asses the deviations from
the idealized assumptions of the $z$ transform.

Figure~\ref{f:outliers} shows the fraction of outliers, $\fout$,
as a function of $\nobs$ for the two light curve types and for the
ZDCF, DCF and evenly sampled ZDCF cases. In all the simulations checked
here, both methods {\em over-estimate} the errors, but the ZDCF
results are always closer than the DCF's to the normal value $\fout=0.3173$.
There is a marked trend of an increase in the over-estimation for
softer power spectra. This is correlated with the deviations from
normality. The flicker noise average kurtosis is ${kurt}=-0.27$ and
the $\chi^{2}$ fit probability for the normality of the marginal
distributions is $P(\chi^{2})\sim0.3$ whereas for the soft power
spectrum the values are ${kurt}=-0.80$ and $P(\chi^{2})\sim0.03$,
respectively. The evenly sampled ZDCF dependence on the underlying
power spectrum is similar to that of the binned ZDCF, confirming that
this effect is not due to the binning procedure.

As anticipated by equation \ref{e:mdt}, the behaviour of the ZDCF,
unlike that of the DCF, is not affected by the number of observations.
The difference between the two methods is most marked in the case
of the flicker noise power spectrum, where the DCF becomes highly
biased as the number of observations decreases.

Qualitatively similar results were obtained for simulations performed
with Gauss-Markov random signals \citet{BH} and signals described
by Gaussian shaped peaks of random height and width, randomly superimposed
on a high order polynomial.

\section{examples}

\label{s:ex}

\subsection{Correlations involving variability time scales}

\label{s:ex1}

The physical origin of AGN variability is poorly understood. One line
of approach to this problem is to look for correlations between the
variability and other AGN properties, such as the luminosity (e.g.
Hook et al. 1994). The following example is a simplified, simulated
version of such an analysis, which was performed on observed data
by \citet{Netzer}. The ZDCF bias depends on the statistical properties
of the light curves, which are generally not known in advance. The
following example demonstrates that the reduced bias of the ZDCF,
even when it cannot be quantified, makes it more efficient in extracting
information from the data than the DCF and thus useful even when applied
to sparsely sampled light curves.

\begin{figure}[t]
\includegraphics[width=0.75\columnwidth]{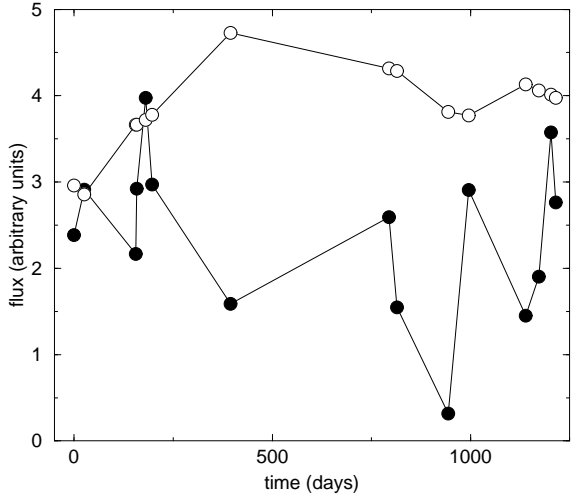}\caption{\label{f:2lc}Two of the simulated light curves used for studying
the correlation between variability timescale and magnitude. Both
light curves are sampled simultaneously 15 times with a clustered
sampling pattern and have a power law index of $\alpha=2$ (open circles)
and $\alpha=1$ (filled circles).}
\end{figure}

The observational situation was simulated as follows. A sample of
30 light curves with power law power spectra, $P(\nu)\propto\nu^{-\alpha}$
were generated as described in section~\ref{s:results}. The 30 light
curves consisted of three light curves for each of the 10 values $\alpha=1.1$,
$1.2$, $\ldots$, $2.0$. All the light curves were then sampled
at the same 15 times (this simulates the situation where the AGN sample
is recorded simultaneously on one photographic plate). The sampling
times (Fig.~\ref{f:2lc}) have a clustered pattern typical of astronomical
observations and were taken from the observed light curve shown in
Fig.~\ref{f:ccf} (The original light curve was diluted down to 15
points by throwing very close observing times). Normally distributed
3\% measurement errors were added to the simulated observations. A
toy model for the magnitude--power spectrum relation was used to associate
each light curve with a magnitude 
\begin{equation}
M(\alpha)=26+\alpha+\epsilon\,,\label{e:mag}
\end{equation}
 where $\epsilon$ is a gaussian variate with zero mean and a standard
deviation of $0.1$. $\epsilon$ represents an additional scatter
in the AGN luminosity that is independent of the power spectrum. This
may be due to measurement errors in $M$ or to a hidden dependence
of $M$ on some physical parameters other than $\alpha$.

The ACF of each of the light curves was estimated by both the DCF
and ZDCF. The 15 observations yielded only two ZDCF bins, and accordingly,
the DCF was also calculated with only two bins. The coherence timescale,
$\to$, is a measure of the typical variability timescale. Following
\citet{Netzer}, $\to$ was formally defined as the shortest lag where
ACF$(\to)=0$ and was estimated by minimal squares fitting of the
straight line $r=1-\tau/\to$ through the DCF and ZDCF points. The
fit took into account both the $\delta r$ and $\delta\tau$ errors.
Finally, Spearman's rank order correlation coefficient between $\to$
and $M$ was calculated for the 30 AGN. Note that the rank order correlation
is insensitive to the exact functional form of $M(\alpha)$, and is
only affected by the ratio between the scatter in $M$ due to the
distribution of $\alpha$ in the sample and that due to $\epsilon$.

\begin{figure}
\begin{tabular}{c}
\includegraphics[width=0.75\columnwidth]{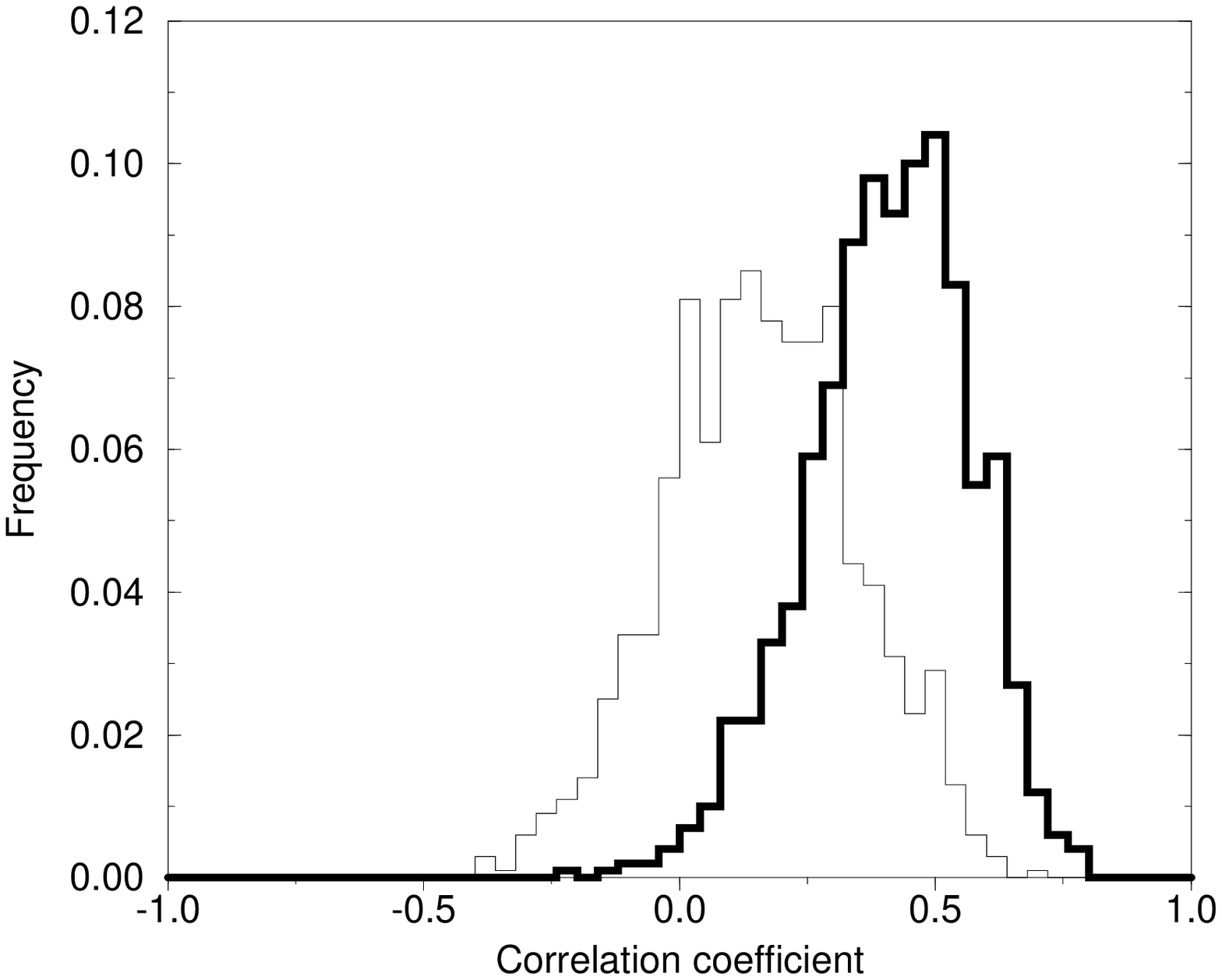}\tabularnewline
\includegraphics[width=0.75\columnwidth]{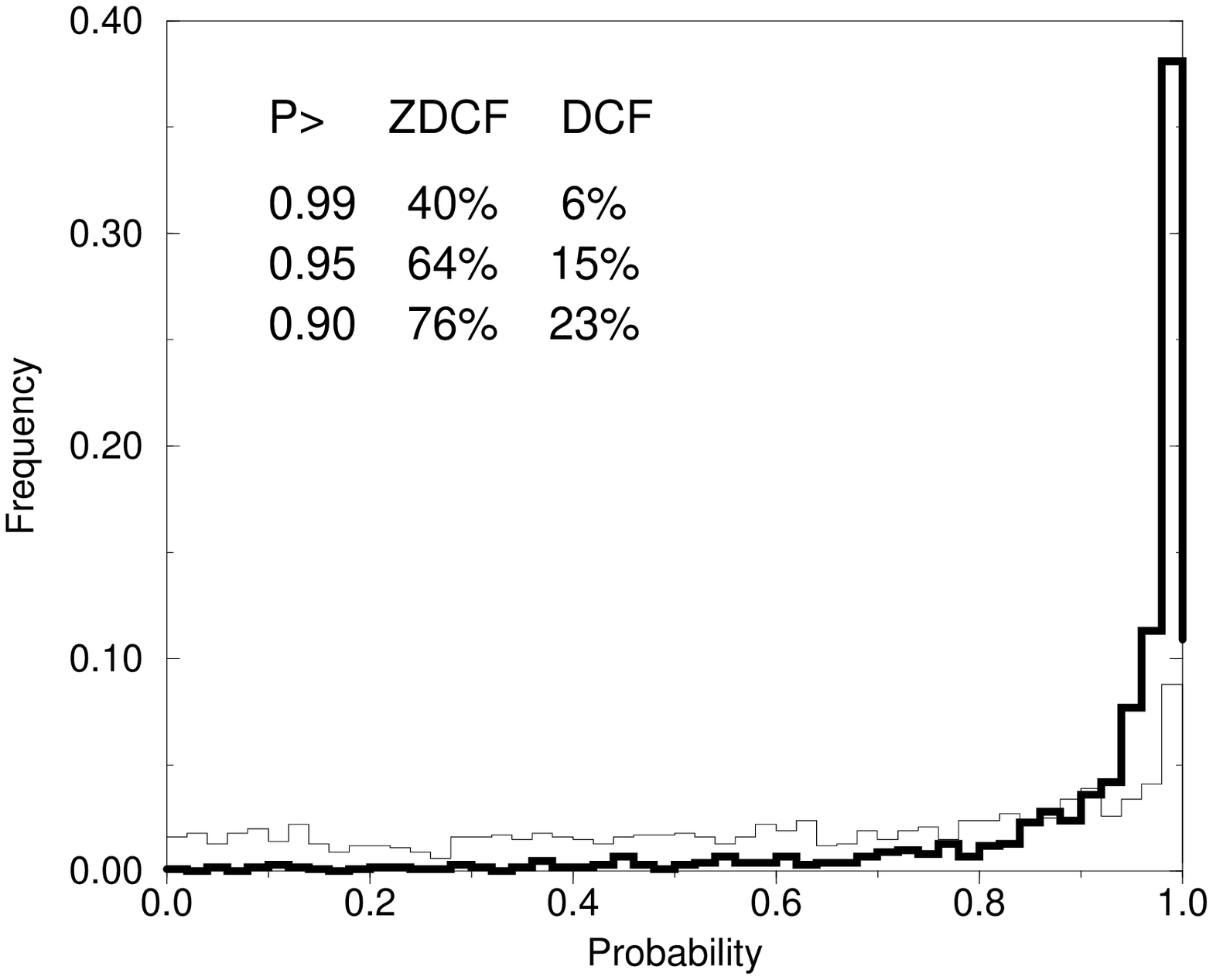}\tabularnewline
\end{tabular}

\caption{\label{f:corr}The sample distribution of Spearman's rank order correlation
coefficient between $\tau_{0}$ and $M$ (top) and its probability
(bottom) calculated with the DCF (thin line) and ZDCF (bold line).
Also listed are the fractions of results with positive correlation
and probabilities greater than 0.90, 0.95 and 0.99.}
\end{figure}

This entire procedure was repeated with $1000$ random samples of
30 light curves each. Fig.~\ref{f:corr} shows the distribution of
the calculated correlation coefficient and of the proability that
the correlation is not random, for both the ZDCF and DCF. For power
law power spectra, $\to$ increases with $\alpha$ and therefore correct
estimates of $\to$ should yield significant positive correlation
coefficients. Fig.~\ref{f:corr} shows that the ZDCF is much more
efficient than the DCF and has a significantly better chance of uncovering
the underlying correlation even in a small sample of sparsely sampled
light curves. For example, in 40\% of the ZDCF simulations, a positive
correlation was detected at the 0.99 significance, as compared to
only 6\% for the DCF. In 76\% of the ZDCF simulations, a positive
correlation was detected at the 0.90 significance, as compared to
only 23\% for the DCF.

\subsection{Time lag estimation by maximum likelihood}

\label{s:ex2}

\begin{figure}[t]
\includegraphics[width=0.75\columnwidth]{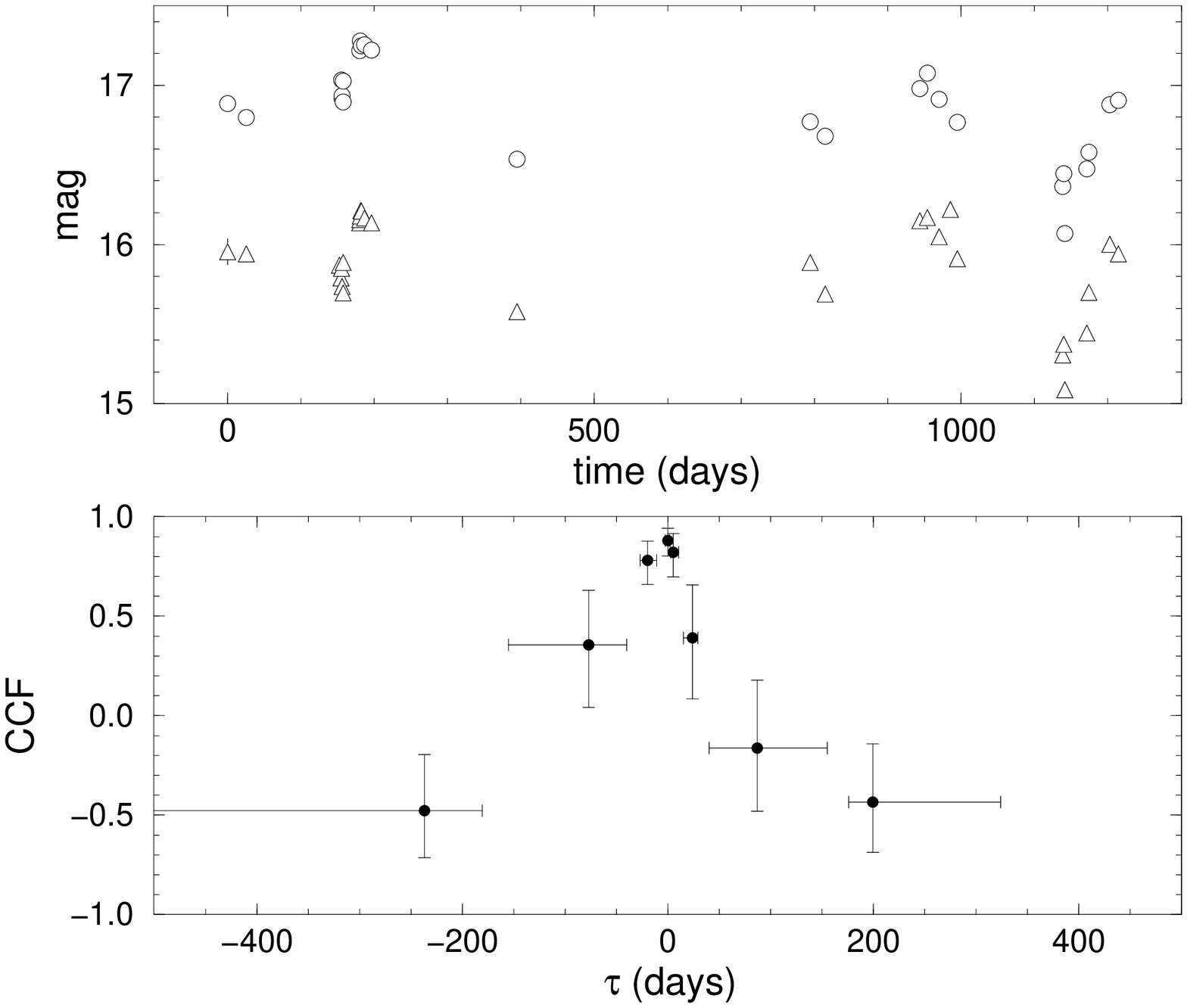} \caption{Top: The R band (triangles) and B band (circles) light curves of the
optically violent variable quasar 3C\,454. The measurement errors
are smaller than the point size. Bottom: The central peak of the ZDCF,
calculated with $\nmin=11$ and 100 Monte Carlo draws for estimating
the effects of the measurement errors.}

\label{f:ccf} 
\end{figure}

Figure~\ref{f:ccf} shows the $R$ and $B$ light curves of the optically
violent variable quasar 3C\,454 \citep{Netzer} and their ZDCF. The
time lag between the two bands, and the question whether it is consistent
with zero, are important for testing models of AGN continuum emission.
In the following example, the ZDCF is used to estimate the uncertainty
in the peak position by using, for the first time in this context,
the fiducial interpretation of the likelihood function. 

The likelihood that point $i$ is the maximum, $L_{i}$, is approximately
the product of the probabilities for point $i$ being larger than
point $j$ 
\begin{eqnarray}
L_{i} & = & \int_{-\infty}^{\infty}d\zeta_{i}\frac{1}{\sqrt{2\pi}s_{z,i}}\exp\left[-\frac{1}{2}\left(\frac{z_{i}-\bar{z}_{i}}{s_{z,i}}\right)^{2}\right]\times\nonumber \\
 &  & \prod_{j\ne i}\int_{-\infty}^{\zeta_{i}}d\zeta_{j}\frac{1}{\sqrt{2\pi}s_{z,j}}\exp\left[-\frac{1}{2}\left(\frac{z_{j}-\bar{z}_{j}}{s_{z,j}}\right)^{2}\right]\,,\label{e:Li0}
\end{eqnarray}
 where $\bar{z}$ and $s_{z}$ are functions of $\zeta$ through $\rho$
and $z$ is a function of $r$, the empirical correlation. The approximation
results from neglecting the possible correlations between the ZDCF
points (e.g. Box \& Jenkins 1970) and from the fact that the ZDCF
points are only approximately normally distributed. It is more convenient
to express $L_{i}$ as an integral over $\rho$ 
\begin{eqnarray}
L_{i} & = & \int_{-1}^{+1}\frac{1}{\sqrt{2\pi}s_{z,i}}\exp\left[-\frac{1}{2}\left(\frac{z_{i}-\bar{z}_{i}}{s_{z,i}}\right)^{2}\right]\times\nonumber \\
 &  & \prod_{j\ne i}\Phi(z_{j},\rho_{i})\frac{d\rho_{i}}{(1-\rho_{i}^{2})}\,,\label{e:Li}
\end{eqnarray}
 where 
\begin{equation}
\Phi(z,\rho)=\int_{-1}^{\rho}\frac{1}{\sqrt{2\pi}s_{z}}\exp\left[-\frac{1}{2}\left(\frac{z-\bar{z}}{s_{z}}\right)^{2}\right]\frac{d\rho^{\prime}}{(1-\rho^{\prime2})}\,.\label{e:phi}
\end{equation}
 $L_{i}$ can be easily calculated numerically. $L_{i}$ reaches its
maximum at the highest ZDCF point, so that the maximum likelihood
(ML) estimate coincides with the ZDCF peak. The sampling distribution
of ML estimators is generally known only in the large sample limit,
where it is Gaussian. Here, having a large sample means having many
pairs of $R$ and $B$ observations of the same continuous light curve,
observed simultaneously by different observers. The actual data set
consists of only one such pair, and it is therefore impossible to
assign a a confidence interval for the peak.

There is an alternative approach for obtaining interval estimates
from the likelihood function, which is related, but not identical,
to Bayesian statistics. The normalized likelihood function, defined
as the {\em fiducial distribution}, is interpreted as expressing
the ``degree of belief'' in the possible value of the estimated
parameter, and the 68\% interval around the likelihood function's
maximum is defined as the 68\% {\em fiducial interval} (e.g. Frodesen,
Skjeggestad \& T{ø}fte 1979). The fiducial function for the peak
is calculated by interpolating between the points of the likelihood
function. The position of the fiducial distribution's maximum is the
maximum likelihood estimate of the time-lag%
\footnote{When the fiducial function is interpolated linearly, the most likely
peak position is that of the highest ZDCF point.%
} and the fiducial interval is that which includes 0.3414 (normal $1\sigma$)
of the area left and right of the maximum, respectively. In practice,
the fiducial interval cannot be narrower than the bin width $\delta\tau_{\pm}(r_{{\rm max}})$.
The fiducial interval can be interpreted as the interval where 68\%
of the likelihood-weighted ensemble of all possible CCFs reach their
peaks. It should be emphasized that a fiducial interval is conceptually
different from a confidence interval, and one should not misinterpret
it as meaning that if the light curves are resampled and their ZDCF
peak calculated, then on the long run, the true peak will lie inside
the fiducial intervals 68\% of the time. A detailed discussion of
the fiducial interval and its relation to Bayesian statistics can
be found in {}{}\citet{KSb}.

The fiducial estimate of the peak has several advantages over other
approaches that were used in AGN variability studies. Unlike methods
that use simulations to estimate the uncertainty on the time-lag \citep[e.g.][]{MN},
the fiducial estimator does not assume anything about the mechanism
that generates the light curves. Another commonly used method for
estimating the peak location is to fit the CCF peak to a peaked function,
such as a parabola or a Gaussian. Some parameterization of the function's
width is then used as a measure of the uncertainty in the peak position.
The arbitrary choice of the function introduces an unwanted degree
of freedom to the final result. It is also often the case that the
peak, unlike the fitted functions, is asymmetric and this leads to
skewed estimates of the peak location. In contrast, The fiducial estimator
does not assume anything about the shape of the peak and thus avoids
these problems.

These advantages come at the price of having to make the approximations
mentioned above in calculating $L_{i}$, as well as the approximation
involved in interpolating the likelihood function%
\footnote{Note that these approximations are also shared by the function-fitting
method, and that methods that use the centroid to parameterize the
time-lag also ignore the biases that may be introduced by the correlations%
}. Another unavoidable consequence of this approach is the assumption
of a Bayesian prior, namely the uniform distribution of the CCF points
in $z$-space.

The CCF peak of 3C\,454 lies between time-lags $\tau=-200$ to $200$
days. The maximum likelihood estimate that is calculated for the points
in this interval is $0.1_{-23.7}^{+10.0}$ days (R lags after B).
This is consistent with the hypothesis that the R and B bands of 3C\,454
vary simultaneously.

\section{Discussion and summary}

\label{s:discuss}

Monitoring light curves of astronomical objects over long periods
of time requires great observational efforts. Inevitable limitations
and difficulties stand in the way of obtaining regular, well sampled
light curves and severely restrict the reliability of spectral analysis.
The result is that in many cases the light curves are analysed directly
in the time domain by the CCF. Such a situation calls for an effort,
on par with the observational one, to develop improved methods for
estimating the CCF, even at the cost of added computational complexity.
In many cases the only way to extract information from meager data
is by modeling. It is therefore important to have error estimates
that can be used for fitting theoretical CCF models to the data and
estimate the location and significance of the CCF extrema.

The ZDCF method attempts to correct the biases that affect the original
DCF. The simulations show that the ZDCF performs better or at least
as good as the DCF under a variety of conditions. The performance
of the ZDCF depends on the ratio between $\mdtt$, the typical time
between observations in a bin, and $\to$, the coherence timescale.
As long as $\mdtt>\to$, the $z$-transform biases are small and the
error estimates are realistic. Since $\mdtt$ does not depend on $\nobs$,
it is possible to increase the time lag resolution of the ZDCF without
increasing its bias by increasing sampling rate (i.e increasing $\nobs$
at a constant total time $T$). When $\mdtt<\to$, the error estimates
of the $z$-transform may be over-estimated. However, in this case
interpolation is justified. This is seen by noting that $\mdtt$ is
a decreasing function of $\tau$, which is approximately bounded from
below by 
\begin{equation}
\mdtt\gtrsim\left(\frac{\nmin-1}{\nmin}\right)\mdt\sim\mdt\,.
\end{equation}
 It then follows that if $\mdtt<\to$, then $\mdt<\to$ and therefore
that the light curves are well sampled. Even in this case the ZDCF
offers a more conservative alternative than interpolation. $\bar{z}$,
unlike $s_{z}$, is relatively unbiased and therefore $\rbin$ estimates
the CCF much better than implied by the formal error estimates. 

The simulations were limited to the uniform random sampling pattern.
Sampling patterns which are far from uniform may introduce complicated
biases to the ZDCF, which can be traced to the fact that the Fourier
transform of the underlying signal is convolved with that of the sampling
pattern itself. This is a fundamental problem that lies beyond the
scope of the ZDCF. A partial solution is to divide the light curve
into segments which are roughly uniformly sampled and perform the
ZDCF on the union of these segments, at the price of estimating the
correlation function only for small values of $\tau$.

To summarize, the calculation of the ZDCF involves the following steps: 
\begin{enumerate}
\item Atypically noisy measurements should preferably be discarded from
the light curve . 
\item All possible pairs of observations, $\{a_{i},b_{j}\}$, are sorted
according to their time-lag $t_{i}-t_{j}$ and binned into equal population
bins of at least 11 pairs. Multiple occurrences of the same point
in a bin are discarded so that each point appears only once per bin
(section~\ref{s:bin}). 
\item Each bin is assigned its mean time-lag and the intervals above and
below the mean that contain $1\sigma$ (normal) of the points each
(section~\ref{s:bin}). 
\item The correlation coefficients of the bins are calculated (equation~\ref{e:rbin})
and $z$-transformed (equation~\ref{e:z}). The error is calculated
in $z$-space (equation~\ref{e:sz}) and transformed back to $r$-space
(equation~\ref{e:dcferr}). 
\item The effect of measurement errors is estimated by Monte Carlo runs,
where at each step a random error is added to each point according
to its quoted error. The ZDCF is averaged in $z$-space and the average
is transformed back to $r$-space (section~\ref{s:error}). 
\end{enumerate}
The resulting CCF error bars are roughly equivalent to normal $1\sigma$
errors. The simulations performed here show a consistent trend towards
over-estimation of the errors and suggest that the error interval
may be as large as $1.4\sigma$ for strongly auto-correlated light
curves. 
 A \textsc{fortran 77} code of the ZDCF method is available from the
author on request.

\appendix

\section{Properties of the binned average}

\label{s:appA}

The binned correlation coefficient is an average of the correlation
coefficient over the bin's time lag interval. The exact nature of
this average can be expressed in terms of the mixture of time-lags
$\{\tau_{i}\}$, associated with the observed pairs $\{a_{i},b_{i}\}$
in the bin. Let the weight $\omega_{i}$ be the fraction of the pairs
with time-lag $\tau_{i}$ (usually $\omega_{i}=1/\nmin$). Define
$\Ema(\tau_{i})$, $\Vma(\tau_{i})$ to be the mean and variance over
the continuous light curve $a$ from $t_{{\rm min}}$ to $t_{{\rm max}}-\tau_{i}$,
$\Epb(\tau_{i})$, $\Vpb(\tau_{i})$ to be the mean and variance over
the continuous light curve $b$ from $t_{{\rm min}}+\tau_{i}$ to
$t_{{\rm max}}$ and $\Cpmab(\tau_{i})$ to be the covariance of the
two light curves with the intervals similarly defined (See Fig.~\ref{f:ranges}).
It is straightforward to show that the binned variance and covariance
over the continuous light curves are 
\begin{eqnarray}
V_{a} & \!\!\!\!= & \!\!\!\!\sum_{i=1}^{n}\omega_{i}\Vma(\tau_{i})+\sum_{i<j}\omega_{i}\omega_{j}(\Ema(\tau_{i})-\Ema(\tau_{j}))^{2}\,,\nonumber \\
V_{b} & \!\!\!\!= & \!\!\!\!\sum_{i=1}^{n}\omega_{i}\Vpb(\tau_{i})+\sum_{i<j}\omega_{i}\omega_{j}(\Epb(\tau_{i})-\Epb(\tau_{j}))^{2}\,,
\end{eqnarray}
 and 
\begin{eqnarray}
C_{a,b}\!\!\!\! & =\!\!\!\! & \sum_{i=1}^{n}\omega_{i}\Cpmab(\tau_{i})+\nonumber \\
 &  & \sum_{i<j}\omega_{i}\omega_{j}(\Ema(\tau_{i})-\Ema(\tau_{j}))(\Epb(\tau_{i})-\Epb(\tau_{j}))\,.
\end{eqnarray}

It is clear that the sample binned correlation coefficient estimates
the quantity 
\begin{equation}
\rho_{{\rm bin}}=\frac{C_{a,b}}{\sqrt{V_{a}V_{b}}}\,,
\end{equation}
 which is not identical to the simple weighted arithmetic mean 
\begin{equation}
\bar{\rho}=\sum_{i=1}^{n}\omega_{i}\frac{\Cpmab(\tau_{i})}{\sqrt{\Vma(\tau_{i})\Vpb(\tau_{i})}}\,,
\end{equation}
 although both values approach each other as the bin width is decreased.
$r$ estimates an averaged value of $\rho$, weighted by the density
distribution of the time-lag points in the bin. In order to reflect
this, the time-lag associated by the ZDCF with the bin is estimated
by the mean lag $\bar{\tau}$. The uncertainty in $\tau$ is estimated
by the intervals $\delta\tau_{\pm}$ that contain 0.3414 (normal $1\sigma$)
of the points in the bin above and below $\bar{\tau}$, respectively.
The resulting error bars are asymmetric and, unlike usual error bars,
do not give the $\pm1\sigma$ uncertainty on the `true' value $\bar{\tau}$
but rather describe the interval over which $\sim2/3$ of the averaging
was performed. Simulations of sparsely sampled light curves, where
$\tau_{{\rm max}}$ is small and $E(\tau_{i})$, $V(\tau_{i})$ and
$C(\tau_{i})$ are therefore calculated on largely overlapping stretches
of the light curves, confirm that $\rho_{{\rm bin}}\simeq\bar{\rho}$
to a very good approximation.

\section{The effect of measurement errors on the DCF}

\label{s:appB}

EK suggest that the sample CCF of the true signals, $r_{x,y}$, can
be recovered from $r_{a,b}$ by a simple analytic correction of replacing
equation~\ref{e:rbin} with 
\begin{equation}
r_{x,y}^{\prime}=\frac{\sum_{i}^{n}(a_{i}-\Ea)(b_{i}-\Eb)/(n-1)}{\sqrt{s_{a}^{2}-s^{2}(n_{a})}\sqrt{s_{b}^{2}-s^{2}(n_{b})}}\,,
\end{equation}
 where $s(n_{a})$ and $s(n_{b})$ are the typical measurement errors.
$|r_{x,y}^{\prime}|>|r_{a,b}|$ by definition, and in the limit of
low S/N, $|r_{x,y}^{\prime}|$ may become arbitrarily large (even
larger than 1), despite the fact that the observed light curves are
almost pure noise. The origin of this problem is that this correction
applies only in the limit of an infinitely sampled, infinite light
curve. This is demonstrated by writing $\rho_{x,x}$ explicitly in
the case of auto-correlation. Assuming for simplicity that the bin
contains only a single time-lag $\tau$, the correlation coefficient
over a continuous finite segment of the light curve is 
\begin{equation}
\rho_{x,x}(\tau)=\frac{C_{a,a}^{\pm}-C_{x,n}^{\pm}-C_{n,x}^{\pm}-C_{n,n}^{\pm}}{\sqrt{(\Vma-V_{n}^{-}-2C_{x,n}^{-})(\Vpa-V_{n}^{+}-2C_{x,n}^{+})}}\,.
\end{equation}
 The errors are assumed to be uncorrelated with the signal and with
themselves, and therefore in the limit of an infinite light curve,
\begin{equation}
C_{x,n}^{\pm}(\tau)=C_{n,x}^{\pm}(\tau)=C_{x,n}^{+}(\tau)=C_{x,n}^{-}(\tau)=0\,,\label{e:cov}
\end{equation}
 and for $\tau\ne0$, 
\begin{equation}
C_{n,n}^{\pm}(\tau)=0
\end{equation}
 also holds. $r_{x,x}^{\prime}$ approaches $\rho_{x,x}$ only in
this limit. Moreover, even in cases where this is an adequate approximation,
it is still unclear whether this estimator has the required statistical
properties (i.e. consistency, efficiency and lack of bias).


\subsubsection*{Acknowledgements}

{I am grateful to Rick Edelson and Julian Krolik for comprehensive
discussions of issues relating to the DCF and ZDCF. Stimulating conversations
with Shai Kaspi, Shai Zucker and Eyal Neistein are much appreciated.
This work was partially supported by the US-Israel Binational Science
Foundation grant no. 89-00179.}

\end{document}